\documentclass[pra,10pt,amsmath,amssymb,aps,twocolumn]{revtex4-1}
\usepackage{comment}
\usepackage{epsfig}
\usepackage{graphicx}
\usepackage{dcolumn}
\usepackage{color,soul}
\usepackage{ulem}
\usepackage{longtable}
\usepackage{placeins}

\usepackage{natbib}

\begin{document}
\title{Anisotropic potential immersed in a dipolar Bose-Einstein condensate}
\author{Neelam Shukla$^1$}
\author{Artem G. Volosniev$^{2,3}$} 
\author{Jeremy R. Armstrong$^1$}

\affiliation{$^1$Department of Physics, University of Nebraska at Kearney, NE-68849 USA}
\affiliation{$^2$Institute of Science and Technology Austria (ISTA), A-KN-3400, Austria}
\affiliation{$^3$Department of Physics and Astronomy, Aarhus University, Ny Munkegade 120, DK-8000 Aarhus C, Denmark
}%

\begin{abstract}
We study a three-dimensional Gross-Pitaevskii equation that describes a static impurity in a dipolar Bose-Einstein condensate (BEC). Our focus is on the interplay between the shape of the impurity and the anisotropy of the medium manifested in the energy and the density of the system.  Without external confinement, properties of the system are derived with basic analytical approaches. For a system in a harmonic trap, 
the model is investigated numerically, using the split-step Crank-Nicolson method. 
 Our results demonstrate that the impurity self-energy is minimized when its shape more closely aligns with the anisotropic character of the bath, in particular a prolate deformed impurity aligned with the direction of the dipoles has the smallest self-energy for a repulsive impurity. Our work complements studies of impurities in Bose gases with zero-range interactions, and paves the way for studies of dipolar polarons with a Gross-Pitaevskii equation.
\end{abstract}

\maketitle
\section{Introduction}
The question of how a system responds to perturbation (stress) is of general importance in physics, in particular, in cold-atom physics. Naturally, isotropic media are the most explored in this context. 
However, progress in realizing 
ultracold dipolar gases brings the perfect playground for going beyond isotropicity~\cite{Chomaz2022}. It becomes possible to investigate the effect of perturbations on quantum degenerate gases that are intrinsically anisotropic.  

In this work, we explore the effect of a static perturbation on an anisotropic medium by placing an impurity at the center of the dipolar Bose gas.
The impurity can be an optically generated potential, which
in principle can be of arbitrary shape in modern cold-atom experiments~\cite{Amico2021}. By calculating the ensuing properties, we gain insight into the differences between dipolar and non-dipolar systems. 

Firstly, we study the Bose gas without external confinement. In this case, the character of the system can be analyzed analytically at least within the local density approximation. We show that even an isotropic impurity leads to an anisotropic response, which reveals the nature of the Bose gas. Further, for an anisotropic impurity, the energy of the system depends strongly on the interplay between the direction of the dipoles and the shape and orientation of the impurity. The dipolar nature of the Bose gas is the most pronounced in the vicinity of the collapse, where the number of bosons displaced by the impurity becomes large. Secondly, we analyze an experimentally relevant system in a harmonic trap. We show that the results derived in the homogeneous setting still hold, advocating for an exploration of static impurities with dipolar gases in a laboratory. 
  
Before proceeding further, we note a connection of our work to a field of mobile impurities in cold gases that has seen a number of advances in the recent past~\cite{schirotzek2009,hu2016,jorgensen2016,desalvo2017, fritsche2021}.
The interactions between the impurity and the medium effectively alter the impurity's properties from its vacuum properties as it is dressed by particles of the medium, turning it into a polaron. Though these are mobile particles, studies of heavy impurities typically lay the ground for understanding the corresponding energy spectra.

The present manuscript is structured as follows: Section~\ref{theory} describes the system and the modified Gross-Pitaevskii equation (GPE) used in the study. Section~\ref{sec:homog_section} presents analytical analysis of a Bose gas without external confinement. Section~\ref{results} contains
numerical simulations of the GPE, along with their discussion. Additionally, we include a study of the time dynamics that follow a sudden immersion of the impurity. This allows us to estimate timescales relevant for the experiment.  A brief summary of the results and an outlook is given in Section~\ref{sec:conclusion}. Finally, we include two appendices with additional information.

\section{Formalism}~\label{theory}

{\it Hamiltonian}. The Hamiltonian that describes the system of interest --  a Bose gas with a static `impurity' potential -- reads 
\begin{multline}
H = \sum_{k=1}^{N} \left[-\frac{\hbar^2}{2m}\frac{\partial^2}{\partial \mathbf{r}_i^2} + V_{trap}(\mathbf{r_i})\right] +  \sum_{i<j}^{N} V_{c}(\mathbf{r}_i - \mathbf{r}_j) \\
+ \sum_{i<j}^{N} V_{d}(\mathbf{r}_i - \mathbf{r}_j) 
+ \sum_{i=1}^{N} V_{i}(\mathbf{r}_i),
\label{Heqn}
\end{multline}
where $m$ is the mass of a boson, and $\mathbf{r}_i$ is the coordinate of the $i$th boson. The bosons are confined by the harmonic-oscillator trapping potential $V_{trap}$. They interact via the contact, $V_c$,  and the dipole-dipole interactions, $V_{d}$, discussed below; the function $V_{i}$ is the impurity-boson potential energy. To differentiate between $V_{trap}$ and $V_{i}$, we note that the latter vanishes at infinity, i.e., $V_{i}(|\mathbf{r}|\to\infty)\to0$.   

To analyze the system, we shall rely on a mean-field approximation. Therefore, we can write the contact interaction as $V_c(\mathbf{r})=4\pi\hbar^2a\delta(\mathbf{r})/m$, where $a$ is the scattering length~\cite{Dalfovo1999,Braaten2006}.
The functional form of the dipole-dipole potential is 
\begin{equation}
V_{d}(\mathbf{r}) = \frac{\mu_0}{4\pi}\frac{\mathbf{d}^2 - 3(\mathbf{d} \cdot \hat{\mathbf{r}})(\mathbf{d} \cdot \hat{\mathbf{r}})}{|\mathbf{r}|^3},
\end{equation}
here $\mathbf{d}$ is the dipole moment of the bosons, $\mu_0$ is the vacuum permeability, and $\hat{\mathbf{r}}$ is the unit vector along the direction of $\mathbf{r}$.
We shall use a system of coordinates in which the $z$-axis is along $\mathbf{d}$, and write the dipole-dipole interaction in a more convenient form
\begin{equation}
V_{d} = \frac{C_{{dd}}}{4\pi}\frac{1-3\mathrm{cos}^2{\theta_{d}}}{r^3},
\end{equation}
where $\theta_{d}$ is the angle between $\mathbf{r}$ and $\mathbf{d}$.   We have adopted the standard notation, $C_{dd}=\mu_0 d^2$, which helps us to introduce a relevant length scale (the `dipolar length') as $a_{{dd}}=C_{{dd}}m/(12\pi \hbar^2)$~\cite{lahaye2009}.
The Hamiltonian in Eq.~(\ref{Heqn}) without $V_{i}$ is well-studied~\cite{lahaye2009,Baranov2012} theoretically. The focus of our paper is on the effect of the impurity.

 {\it Gross-Pitaevskii equation.} To analyze the system, we rely on a mean-field ansatz, i.e., we assume that the ground state of the Hamiltonian can be approximated by a product state $\psi(\mathbf{r}_1)\psi(\mathbf{r}_2)...\psi(\mathbf{r}_N)$ well. To find the function $\psi$, we solve the GPE
\begin{multline}
 - \frac{\hbar^2}{2m}\frac{\partial^2\psi(\mathbf{r})}{\partial \mathbf{r}^2} +  V_{trap}(\mathbf{r}) \psi(\mathbf{r}) +gN|\psi(\mathbf{r})|^2\psi(\mathbf{r}) \\
 +N \int \mathrm{d}\mathbf{x} V_{d}( \mathbf{r}-\mathbf{x})|\psi(\mathbf{x})|^2 \psi(\mathbf{r})  = \left(\mu-V_{i}(\mathbf{r})\right) \psi(\mathbf{r}),
 \label{eq:GPE}
\end{multline}
where $g=4\pi\hbar^2 a/m$ reflects the strength of the contact interaction  and
$\mu$ is the chemical potential. Any solution to the GPE is subject to the normalization condition $\int\mathrm{d}\mathbf{r} |\psi(\mathbf{r})|^2=1$.

In the next section, we present various approximations for analytical analysis of Eq.~(\ref{eq:GPE}) in a homogeneous case. In Sec.~\ref{results},  we calculate the energies and densities in a trapped case. To this end, we solve the GPE using the split-step Crank-Nicolson method. Our numerical code is a modification of an open-source software described in Ref.~\cite{kumar2015}.  

\section{Homogeneous system}
\label{sec:homog_section}

{\it The Thomas-Fermi limit.} We start by considering an infinite medium with $V_{trap}=0$. We assume that the potential $V_i$ changes `weakly and slowly' (see the discussion below for a weak impurity potential) so that we can employ the Thomas-Fermi approximation and solve the equation
\begin{equation}
V_{i}(\mathbf{r})+g \delta n(\mathbf{r})=
 -\int \mathrm{d}\mathbf{x} V_{d}( \mathbf{r}-\mathbf{x} )\delta n(\mathbf{x}),
 \label{eq:GPE_Thomas_Fermi}
\end{equation}
where $\delta n(\mathbf{x})=N|\psi(\mathbf{x})|^2-n_0$ determines  the density of the Bose gas in the presence of the impurity, $n_0=N/\mathcal{V}$ is the average density; $n_0=N \int\mathrm{d}\mathbf{r}|\psi(\mathbf{r})|^2/\mathcal{V}$ with $\mathcal{V}$ being the volume of the system. $n_0$ is also
the density far away from the impurity because $\delta n(\mathbf{r}\to\infty)\to0$ as we will demonstrate below.
This implies that $n_0$  fixes the chemical potential, i.e., $\mu=gn_0$ (it is independent of dipole-dipole interactions in a homogeneous setting)~\footnote{ 
One could have also used the chemical potential $\mu=g n$, where $n$ is the density of the Bose gas without the impurity. As $n$ coincides with $n_0$ in the limit $\mathcal{V}\to\infty$, this does not affect our results.  
We illustrate the fact that $n=n_0+O(1/\mathcal{V})$ here for a non-dipolar Bose gas. If $V_d=0$, the Thomas-Fermi approximation reads $g\delta n(\mathbf{x})=g (n-n_0) -V_i(\mathbf{x})$. The normalization condition demands that $\int\mathrm{d}\mathbf{r}\delta n(\mathbf{r})=0$, which implies that $n=n_0+\int \mathbf{d}\mathbf{r}V_i(\mathbf{r})/(g\mathcal{V})$. By assumption, $V_i$ has finite support, so that $\int \mathbf{d}\mathbf{r}V_i(\mathbf{r})/(g\mathcal{V})$ is indeed of the form $O(1/\mathcal{V})$. 
}.
 In this section, we shall use $n_0^{-1/3}$ as the unit of length, and $\mu$ as the unit of energy for presenting our findings in a dimensionless form.

Equation~(\ref{eq:GPE_Thomas_Fermi}) has a non-local character, i.e., the density at any given position is determined in part by the density of its surrounding neighborhood. This makes it hard, if not impossible, to solve the equation in real space explicitly for a general form of the impurity potential (though see ~\cite{Eberlein2005} for solutions in some special cases).  
At the same time, linearity of Eq.~(\ref{eq:GPE_Thomas_Fermi}) with respect to $\delta n$ allows us to find its Fourier transform $
\delta \tilde n(\mathbf{k})=\int \mathrm{d}\mathbf{r}\delta n(\mathbf{r})e^{-i\mathbf{k}\cdot\mathbf{r}}$ easily:
\begin{equation}
\delta \tilde n(\mathbf{k})=-\frac{\tilde V_i(\mathbf{k})}{g+\tilde V_d(\mathbf{k})},
\label{eq:delta_n_k_hom}
\end{equation}
where $\tilde V_i$ and $\tilde V_d$ are Fourier transforms of the impurity and dipole-dipole potentials, respectively.  
In the case of the dipole potential, $\tilde V_d(\mathbf{k})=C_{dd}(\cos^2\alpha-1/3)$~\cite{Goral2000,lahaye2009} with $\alpha$ being the angle between $\mathbf{d}$ and $\mathbf{k}$. This expression can be conveniently re-written using the second Legendre polynomial:  $\tilde V_d(\mathbf{k})=2C_{dd}P_2(\cos(\alpha))/3$.

The density in real space can be written in integral form as:
\begin{equation}
\delta n(\mathbf{r})=-\frac{1}{g(2\pi)^3}\int\mathrm{d}\mathbf{k} \frac{\tilde V_i(\mathbf{k})}{1+2\varepsilon_{dd} P_2(\cos\alpha)} e^{i\mathbf{k}\cdot\mathbf{r}},
\label{eq7}
\end{equation}
where $\varepsilon_{dd}=a_{dd}/a$ is a dimensionless ratio that determines the relative importance of dipolar physics. In particular, the system is unstable against a collapse if $\varepsilon_{dd}$ is larger or equal to one~\cite{lahaye2009}. Notice that the density is anisotropic even if the impurity potential is isotropic. 

To illustrate this, let us consider $\varepsilon_{dd}\to0$ (weak dipole-dipole interactions). Using the plane-wave expansion [$e^{i\mathbf{k}\cdot\mathbf{r}}=4\pi\sum_{l,m}i^lj_l(kr)Y_{l}^m( \mathbf{\hat k})(Y_{l}^m( \mathbf{\hat r}))^*$ with $Y_l^m$ being the spherical harmonics], we derive
\begin{equation}
\delta n(\mathbf{r})\simeq -\frac{V_i(\mathbf{r})}{g}-\frac{\varepsilon_{dd}P_2(\cos\theta)}{g\pi^2}\int \mathrm{d}k k^2 \tilde V_i(k)j_2(kr),
\label{eq:delta_n_weak}
\end{equation}
where $j_l$ is a spherical Bessel function, and $\theta$ is the polar angle of $\mathbf{r}$.
The first term here is the Thomas-Fermi profile for a non-dipolar gas~\cite{Dalfovo1999}. The second part is due to the dipolar character of the medium; we have used the isotropicity of $V_i$ to simplify it. 

\begin{figure}
\centering
\includegraphics[width=0.4\textwidth]{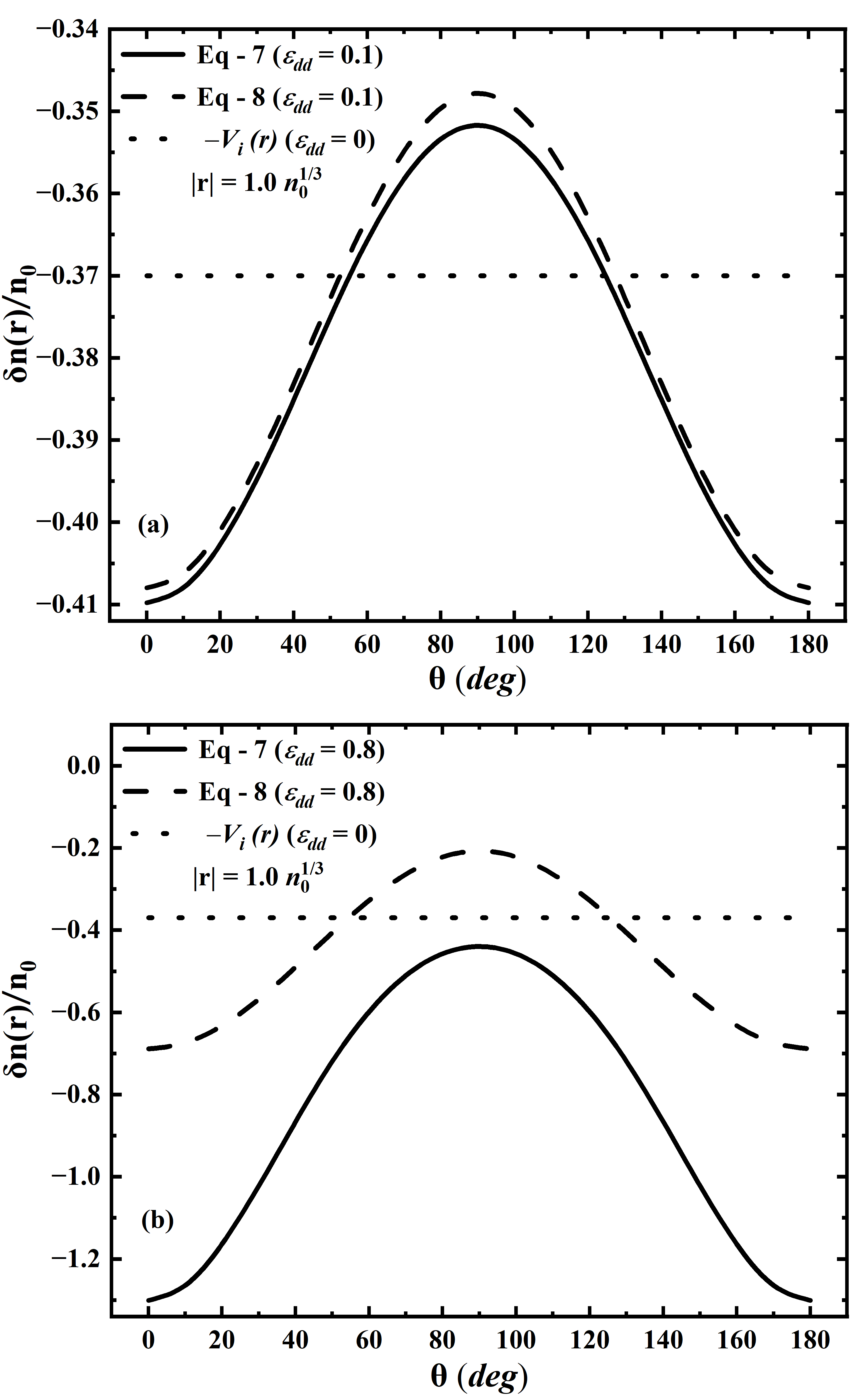}
\caption{Comparison of the results obtained from Equations~(\ref{eq7}) and~(\ref{eq:delta_n_weak}) at different values of $\varepsilon_{dd}$, plotted with respect to the polar angle ($\theta$), i.e., $\mathbf{r}=|\mathbf{r}|(\sin(\theta),0,\cos(\theta))$. The value of the impurity potential is shown with a horizontal dotted line. Note that panel (b) makes it clear that  Eq.~(\ref{eq7}) is not valid for large values of $\varepsilon_{dd}$ as $\delta n/n_0+1$ must be  non-negative (recall that $\delta n+n_0=N|\psi|^2$). }
\label{fig:1_0}
\end{figure}

The uncomplicated form of Eq.~(\ref{eq:delta_n_weak}) will be modified for larger values of $\varepsilon_{dd}$. In particular, the density will include higher order harmonics, $P_{2n}(\cos\theta)$.
We clarify this in Fig.~\ref{fig:1_0} by comparing Eqs.~(\ref{eq:delta_n_weak}) and~(\ref{eq7}) for a strong spherical Gaussian impurity, $V_i(\mathbf{r})=\mu \exp(-n_0^{2/3}r^2)$,
and different values of $\varepsilon_{dd}$.  At small values of $\varepsilon_{dd}$, Eqs.~(\ref{eq7}) and~(\ref{eq:delta_n_weak}) agree with each other, while becoming more different as the dipolar interaction becomes stronger. It is seen clearly that the Thomas-Fermi approach approximation has broken down in the bottom panel of Fig.~\ref{fig:1_0} as it produces a non-physical density at small ($\theta\lesssim \pi/6$) and large ($\theta\gtrsim 5\pi/6$) angles.  This is discussed in more detail at the end of this section.  These two equations do, however, retain similar angular dependence at all values of $\varepsilon_{dd}$. From Fig.~\ref{fig:1_0}~a), where the results agree also quantitatively, we conclude that the density increases in comparison to the non-dipolar solution, $\varepsilon_{dd}=0$, in the direction  perpendicular to $z$, i.e., in the $xy$ plane. Indeed, it is easier to deform the system in this direction because the corresponding phononic excitations are softer~\cite{lahaye2009}.

To further highlight the anisotropic character of Eq.~(\ref{eq:delta_n_weak}), let us study the gradient of the density. In the contact case, we have $\nabla n= \mathbf{F}/g$, where the `force' is given by $\mathbf{F}=-\nabla V_i$. This simple law follows from conservation of chemical potential throughout the sample and the fact that the medium is isotropic, so that we cannot assign any tensor to it. For the dipolar medium the situation is different, and the gradient is given not only by the external force, but also by properties of the medium (e.g., internal strain), which are anisotropic and non-local.

{\it Self-energy of the impurity}. It is worth noting that Eq.~(\ref{eq:GPE}) with the reduced mass instead of $m$
can describe also a mobile impurity just as in a non-dipolar case~\cite{gross1962,Volosniev2017,Hryhorchak_2020,Jager2020,Enss2020,Guenther2021}. Therefore, even though, our work focuses on the effect of static impurities,
our results could also be useful for mobile impurities. Below, we 
use the densities in the Thomas-Fermi limit to calculate two paradigmatic properties typically studied for mobile impurities. First, we compute the self-energy of the impurity defined as the difference of the energies of the system with and without the impurity assuming a fixed chemical potential:
\begin{equation}
 E_{\mathrm{self}}=E[V_i]-E[V_i=0],
 \label{eq:self_energy_def}
\end{equation}
where $E[V_i]$ is the energy of the system with the impurity potential $V_i$. This energy is one of the key characteristics for systems with impurities. The self-energy can be thought of as the chemical potential of the impurity or the change in energy of the bath due to the presence of the impurity~\cite{scazza2022}.  Within our approach,  $E_{\mathrm{self}}$ has the form (see App.~\ref{app:Eq13}):
\begin{align}
 E_{\mathrm{self}}=-\frac{gN^2}{2}&\int\psi(\mathbf{x})^4\mathrm{d}\mathbf{x}+
\frac{g\mathcal{V}n_0^2}{2}- \nonumber \\
 &\frac{N^2}{2}\int\psi(\mathbf{x})^2V_d(\mathbf{r}-\mathbf{x})\psi(\mathbf{r})^2\mathrm{d}\mathbf{r}\mathrm{d}\mathbf{x},
 \label{eq:self_energy_general}
\end{align}
where $\mathcal{V}$ is the volume of the system. After straightforward calculations, we write this expression as 
\begin{equation}
E_{\mathrm{self}}=E_{\mathrm{self}}^0-\frac{1}{2g}\int \delta n(\mathbf{r})V_{d}(\mathbf{x}-\mathbf{r})V_i(\mathbf{x})\mathrm{d}\mathbf{r}\mathrm{d}\mathbf{x},
\label{eq:self_TF}
\end{equation}
where 
$E_{\mathrm{self}}^0=n_0\int V_i(\mathbf{r})\mathrm{d}\mathbf{r}-\frac{1}{2g}\int V_i^2(\mathbf{r})\mathrm{d}\mathbf{r}$ is the contribution to the energy that is independent of the dipolar nature of the Bose gas. Let us assume that the dipole-dipole interactions are weak (i.e., $\varepsilon_{dd}\to0$), then we can estimate the dipolar contribution to the self-energy as
\begin{equation} 
E_{\mathrm{self}}-E_{\mathrm{self}}^0\simeq \frac{1}{2g^2}\int V_i(\mathbf{r})V_{d}(\mathbf{x}-\mathbf{r})V_i(\mathbf{x})\mathrm{d}\mathbf{r}\mathrm{d}\mathbf{x}.
\label{eq:self_energy_TF}
\end{equation}
This equation clearly shows that the self-energy depends not only on the anisotropy of the dipole-dipole interactions, but also on the anisotropy of the impurity potential. Indeed, if the potential $V_i$ is elongated along the $z$-direction, then the right-hand-side of Eq.~(\ref{eq:self_energy_TF}) is negative. The opposite is true for the impurity potentials elongated along the $x$-axis.  We shall illustrate this dependence numerically in the next section for a trapped system. 

{\it Number of bosons in a  `dressing' cloud}. The number of bosons directly affected by the presence of the impurity can be estimated as follows ($\delta N= \int \mathrm{d}\mathbf{r}\delta n(\mathbf{r})$; see~App.~\ref{app:Eq13}) 
\begin{equation}
 \delta N=-\frac{\int \mathrm{d}\mathbf{r} V_i(\mathbf{r})}{g\sqrt{3\varepsilon_{dd}(1-\varepsilon_{dd})}}\mathrm{tan}^{-1}\left[\sqrt{\frac{3\varepsilon_{dd}}{1-\varepsilon_{dd}}}\right].
 \label{deltaN}
\end{equation}
This expression shows that $|\delta N|$ is an increasing function of $\varepsilon_{dd}$, implying that the dipolar medium leads to a `more heavy' dress of the impurity. This can be most easily visualized by considering limiting cases. For weak dipole-dipole interactions, $\varepsilon_{dd}\to0$, we derive 
\begin{equation}
\delta N\simeq -\frac{1}{g}\int \mathrm{d}\mathbf{r} V_i(\mathbf{r})\left(1+\frac{4\varepsilon_{dd}^2}{5}\right).
\end{equation} 
Note that this expression depends on the square of $\varepsilon_{dd}$, because the linear contribution from Eq.~(\ref{eq:delta_n_weak}) averages out to zero. In the vicinity of the collapse, $\varepsilon_{dd}\to1$, 
\begin{equation}
\delta N\simeq -\frac{\pi \int \mathrm{d}\mathbf{r} V_i(\mathbf{r})}{2\sqrt{3}g\sqrt{1-\varepsilon_{dd}}}.
\end{equation}
This result is likely beyond the limits of applicability of local-density approximation, see below. Still, it shows that the number of bosons in the dressing cloud becomes large close to collapse. For a mobile impurity, this should lead to strong renormalization of the effective mass in agreement with previous studies of dipolar Bose polarons~\cite{Kain2014,Ardila2018,Volosniev2023}. The mass renormalization is however direction-dependent with the strongest effect in the $xy$-plane, which cannot captured by the angle-averaged quantity $\delta N$.

{\it Limits of applicability of the Thomas-Fermi approximation}. The Thomas-Fermi approximation is valid whenever the quantum pressure term, proportional to $\frac{\partial^2\psi(\mathbf{r})}{\partial \mathbf{r}^2}$ can be neglected in Eq.~(\ref{eq:GPE}). For a non-dipolar gas that can be done when the impurity potential changes weakly on length scales given by the local healing length of the gas, $\xi\sim 1/\sqrt{g n(\mathbf{r})}$. Obviously the Thomas-Fermi approximation fails when the impurity potential is so strong that $n(\mathbf{r})\to 0$, which happens for example in trapped gases close to the edges of the condensate~\cite{Dalfovo1999}. For a dipolar gas, one can generalize the condition above assuming that the healing length depends on the direction.

To illustrate this, let us  study a weak impurity potential, namely, we consider Eq.~(\ref{eq:GPE}) without a trap and $V_i\to0$. We look for a solution $\psi$ in the following form (cf.~Ref.~\cite{Branko2013}): 
\begin{equation}
\psi(\mathbf{r}) \simeq \frac{1+ f(\mathbf{r})}{\sqrt \mathcal{V}}.
\end{equation}
We assume that $f$ is small, and satisfies the equation  
\begin{equation}
    \frac{\hbar^2}{2m}\nabla^2 f(\mathbf{r}) = 2\mu f+\frac{2\mu}{g}
    \int\mathrm{d}\mathbf{x}V_d(\mathbf{r}-\mathbf{x})f(\mathbf{x})+V_i,
\end{equation}
which allows us to find Fourier transform $\tilde f(\mathbf{k})$ as
\begin{equation}
\tilde f(\mathbf{k})=-\frac{\tilde V_i(\mathbf{k})}{\frac{\hbar^2k^2}{2m}+2\mu+2\frac{\mu}{g}\tilde V_d(\mathbf{k})}.
\end{equation}
By comparing this equation to Eq.~(\ref{eq:delta_n_k_hom}), we conclude that the Thomas-Fermi approximation is valid if the characteristic momenta of $\tilde V_i(\mathbf{k})$ are much smaller than $\sqrt{4\mu m/\hbar^2}\sqrt{1+2\varepsilon_{dd} P_2(\cos\alpha)}$.
For $\varepsilon_{dd}=0$, this condition implies that the potential must change weakly on length scales given by the healing length, validating our assertion above. For a dipolar gas,
this condition depends strongly on $\varepsilon_{dd}$ and on the direction. 
In particular, for $\varepsilon_{dd}$ close to one, $1+2\varepsilon_{dd} P_2(0)$ is small and the Thomas-Fermi approximation should be used with great care.

Finally, we mention another indication that the local-density approximation fails,
which follows from the observation that $\delta n(\mathbf{r})/n+1$ should be non-negative. Assuming repulsive spherically symmetric impurity with its maximum at the origin, the fulfillment of this condition can be checked by considering $\delta n(0)/n+1$, which using Eq.~(\ref{eq7}) leads to 
\begin{equation}
1-\frac{\int \mathrm{d}k \tilde V_i(k) k^2}{2\pi^2 \mu\sqrt{3\varepsilon_{dd}(1-\varepsilon_{dd})}}\mathrm{tan}^{-1}\left[\sqrt{\frac{3\varepsilon_{dd}}{1-\varepsilon_{dd}}}\right]\geq 0.
\end{equation}
This condition clearly fails when the impurity potential is too strong, i.e., $V_i(0) \gtrapprox \mu$, or when the system is close to the dipolar collapse, $\varepsilon_{dd}\to 1$.  In both of these cases, the corresponding healing lengths diverge undermining the applicability of the Thomas-Fermi approximation.
We illustrate this in Fig.~\ref{fig:1_0} where a strong impurity potential and a relatively large value of $\varepsilon_{dd}$ lead to $\delta n/n_0+1<0$ signalling the breakdown of the Thomas-Fermi approximation.

\begin{figure}
\centering
\includegraphics[width=0.4\textwidth]{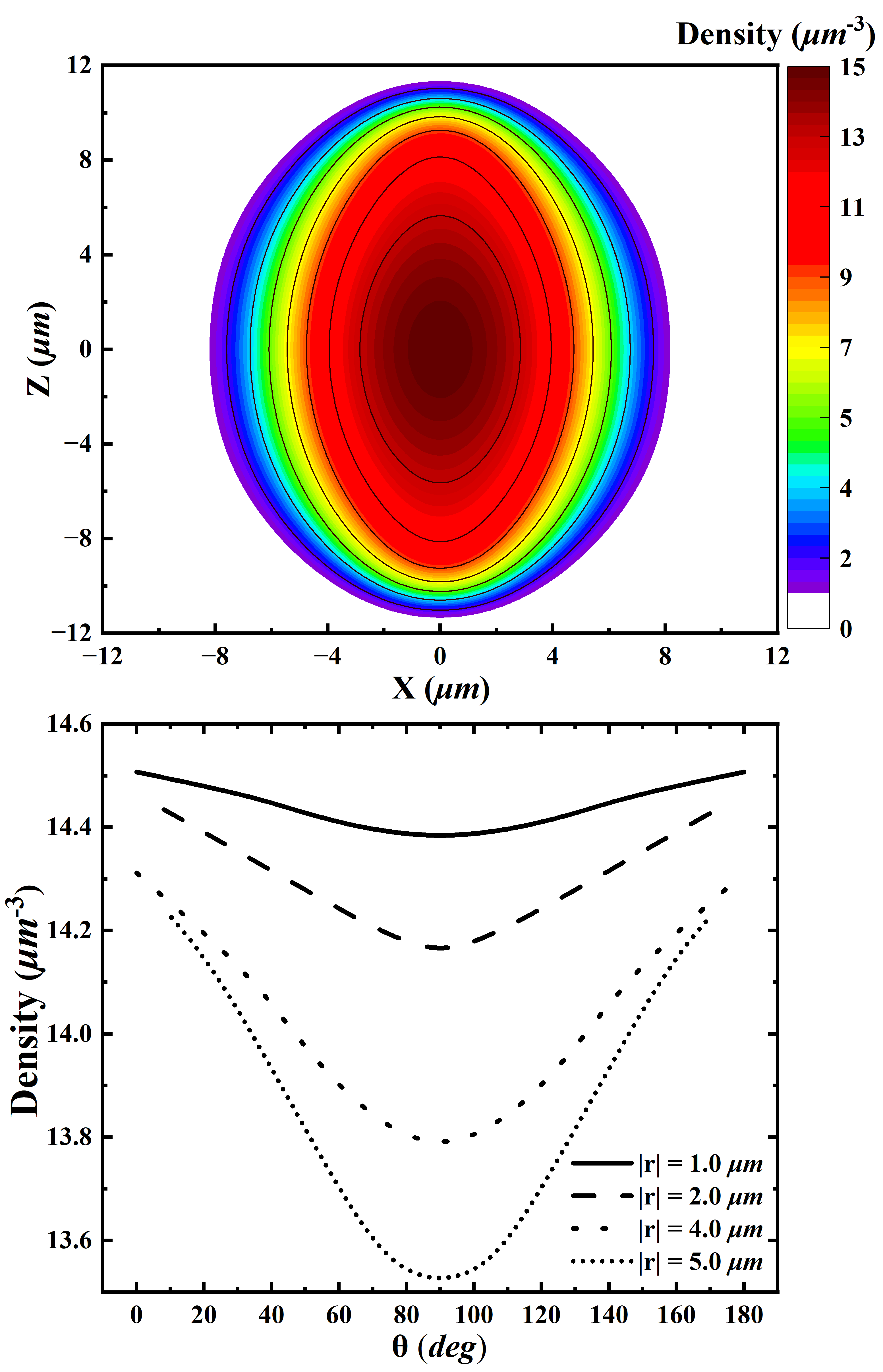}
\caption{Density distribution of the BEC in the absence of the impurity potential. Panel (a) shows the density in the \textit{xz}-plane. Panel (b) shows the density for a given value of $|\mathbf{r}|$ as a function of the polar angle. 
}
\label{fig:1}
\end{figure}

\section{Trapped system}~\label{results}
Guided by the results above, we now conduct  a detailed analysis across a range of scenarios, which should help to identify the physics described in the previous section in a laboratory. To this end,
we consider a harmonically trapped system of dysprosium atoms whose large dipolar length \(a_{{dd}} \approx 130a_0\) puts them at the forefront of current experimental studies~\cite{Chomaz2022}.
For simplicity, the trapping potential is assumed to be symmetric, with the form given by:
\begin{equation}
V_{\text{trap}} = \frac{1}{2} m \omega^2 r^2,
\end{equation}
where \(\omega\) is the trapping frequency.
The corresponding harmonic oscillator length is $L=\sqrt{\hbar/(m\omega)}$.
In our numerical simulations, we use
\(L = 1 \mu\text{m}\), which is representative of experimentally relevant values~\cite{Lu2011}.
We set the number of atoms to \(N = 2 \times 10^4\).  
The atomic scattering length is chosen to be  \(a = 150a_0\), which makes the gas stable  ($a>a_{dd}$). This $a$ is close to the background value in $^{162}$Dy samples, otherwise our choice 
is rather arbitrary as $a$ can be tuned using external magnetic fields, see, e.g.,~Ref.~\cite{Chomaz2022}. 

In Fig. \ref{fig:1}, the density distribution of the dipolar gas in the absence of the impurity is shown in the \textit{xz}-plane, parallel to the plane of polarization. As expected~\cite{Yi2000,Santos2000}, the graph demonstrates that the intrinsic interactions between the dipoles of the system make the system elongated along the \textit{z}-axis, the direction in which the dipoles are polarized. We stress that the trap itself is isotropic, so that the anisotropy observed in the density is purely from the interactions within the system (cf.~Sec.~\ref{sec:homog_section}). For the considered parameters of the Bose gas, this anisotropy is weak close to the origin (see, e.g., the almost flat curve at $|\mathbf{r}|=1\mu$m in Fig. \ref{fig:1}b)),
providing a suitable testbed for investigating the anisotropy induced by the impurity.  
\subsection{Introducing the Impurity}

 We assume an impurity described by the potential
\begin{equation}
V_{i}(x,y,z) = V_0\exp\left[ -\left( \frac{x^2 + y^2}{a^2L^2} + \frac{z^2}{b^2L^2} \right) \right].
\label{ImpEqn}
\end{equation}
Here, $a$ and $b$ parameterize the width of the impurity potential in the transverse and longitudinal directions. The parameter $V_0$ determines the strength of the impurity potential. This parameter is positive in main part of this study, i.e., the impurity repels the bosons, see App.~\ref{app:attractive}
for a brief discussion of the case with $V_0<0$.
It was set to be $5 \hbar\omega$ in our calculations, implying that the effect of the impurity is strong (non-perturbative), i.e., a number of collective modes can be excited in the Bose gas.
The potential in Eq.~(\ref{ImpEqn}) can be optically generated in a cold Bose gas~\cite{Amico2021}, which would provide a physical realization of the impurity studied here in a laboratory~\footnote{Relevant examples might be dimple~\cite{Andrews1997,StamperKurn1998,weber2003,Ma2004,comparat2006,Stellmer2013} and `bubble' traps~\cite{Carollo2022}
}.

\begin{figure}
    \centering
    \includegraphics[width=0.4\textwidth]{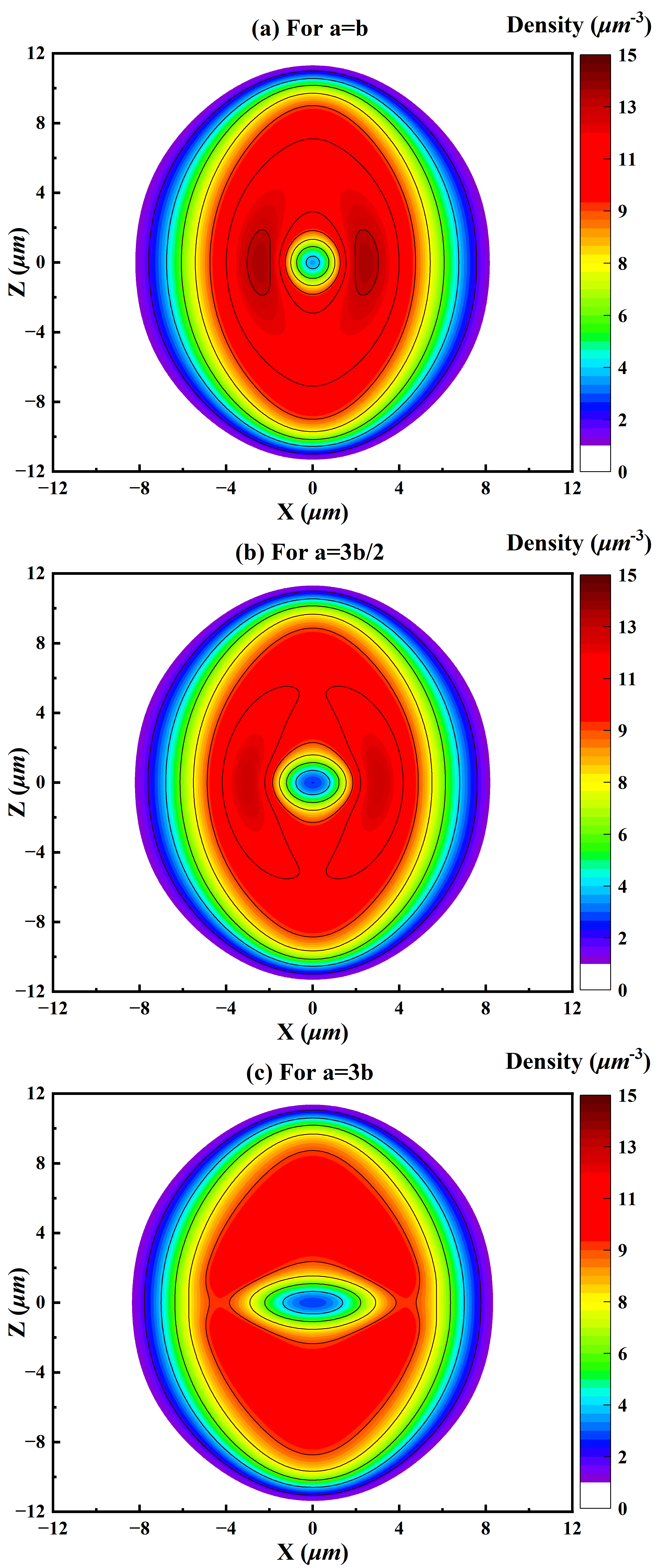}
    \caption{Density of the system (BEC + static impurity) in the $xz$-plane for different deformation ratios $a/b$, cf. Eq.~(\ref{ImpEqn}).
    }
    \label{fig:2}
\end{figure}

First, to understand the effect of the static impurity on the BEC, we calculate the density of the Bose gas for various deformation ratios. Here, the ``deformation ratio'' refers to a systematic variation in the values of \(a\) and \(b\) in Eq.~(\ref{ImpEqn}), while keeping the volume of the impurity ($4\pi a^2bL^3/3$, treating it as an ellipsoid) constant at 1 $\mu$m$^3$.  The volume integral of the potential, $\int\mathrm{d}\mathbf{r}V_i(\mathbf{r})$, is then also constant at $3V_0L^3\sqrt{\pi}/4$.  
We showed in Sec.~\ref{sec:homog_section} that the volume integral of the impurity is a crucial property of the potential that determines the energy of the system in the homogeneous case. Therefore, by fixing it we can more directly see the effect of the dipolar nature of the bath, i.e., we can isolate the effect of the last term in Eq.~(\ref{eq:self_TF}). [Note, however, that the interplay between the external trap and the impurity potential also plays an important role in the results presented in this section.] For the sake of simplicity, here we will mostly discuss results for \(a=b\), \(a=3b/2\), and \(a=3b\).

Figure~\ref{fig:2}(a) illustrates the scenario of an isotropic impurity (\(a=b\)), as evidenced by the circular contours at the origin.
The emergence of two central peaks shows the effect of the static impurity repelling the dipoles away from the origin. The peaks of the densities are located along the $z$=0 lines in agreement with the discussion in Sec.~\ref{sec:homog_section} (see in particular Fig.~\ref{fig:1_0}). This figure shows a 2D cut of a 3D system, so one should bear in mind that instead of actual peaks in the density, the highest density exists in a torus around the $z$-axis centered in the $xy$-plane.  Further away from the impurity, the density is dominated by the harmonic trap (cf. Fig.~\ref{fig:1}).

Figures~\ref{fig:2}(b)-(c) display results for anisotropic impurity potentials  (\(a \neq b\)), with the impurity elongated along the \textit{x}-axis, which induces a competition between intrinsic anisotropy of the system and anisotropy of the impurity. The elongation of the impurity dominates the properties of the dipolar gas at the origin. In particular, we see that contour plots of the density are elongated predominantly along the \textit{x}-axis in the vicinity of the origin.

In Fig.~\ref{fig:2}(c), which represents the maximum deformation analyzed in this study, the peak density regions now appear above and below the $xy$-plane, though they appear much broader and diffuse than the other cases.  The large deformation of the impurity has forced the system to reconstitute itself away from the $xy$-plane and assemble in large caps separated along the dipolar axis.  While the behavior far from the impurity remains similar to the previous cases, the system's response in the region of the impurity is clearly dependent on the precise shape and orientation of the impurity, the latter property we explore in the next section.

\subsection{Orientation-Dependent Effects of Impurity}
To further analyze the interplay between the anisotropy of the impurity and the dipolar nature of the bath, in particular, the restoring of the dipolar symmetry far from the impurity,  
 we examine the effect of rotating the impurity about the \textit{y}-axis. To achieve this, we employ a rotation matrix to change the functional form of the impurity potential in Eq.~(\ref{ImpEqn}). As a result, we obtain:
\begin{multline}
    V_{i} = V_0\exp\left[-\left(\frac{(x\cos\varphi + z\sin\varphi)^2 + y^2}{a^2L^2}\right.\right.\\
    \left.\left. + \frac{(z\cos\varphi - x\sin\varphi)^2}{b^2L^2}\right)\right],
    \label{eq:potential_rotated}
\end{multline}
where, the angle $\varphi$ is defined relative to the $x$-axis. Neither the volume of the impurity ($4\pi a^2bL^3/3 = 1 \mu$m$^3$) nor the volume integral of the potential are changed by rotation. 

\begin{figure}
    \centering
    \includegraphics[width=0.4\textwidth]{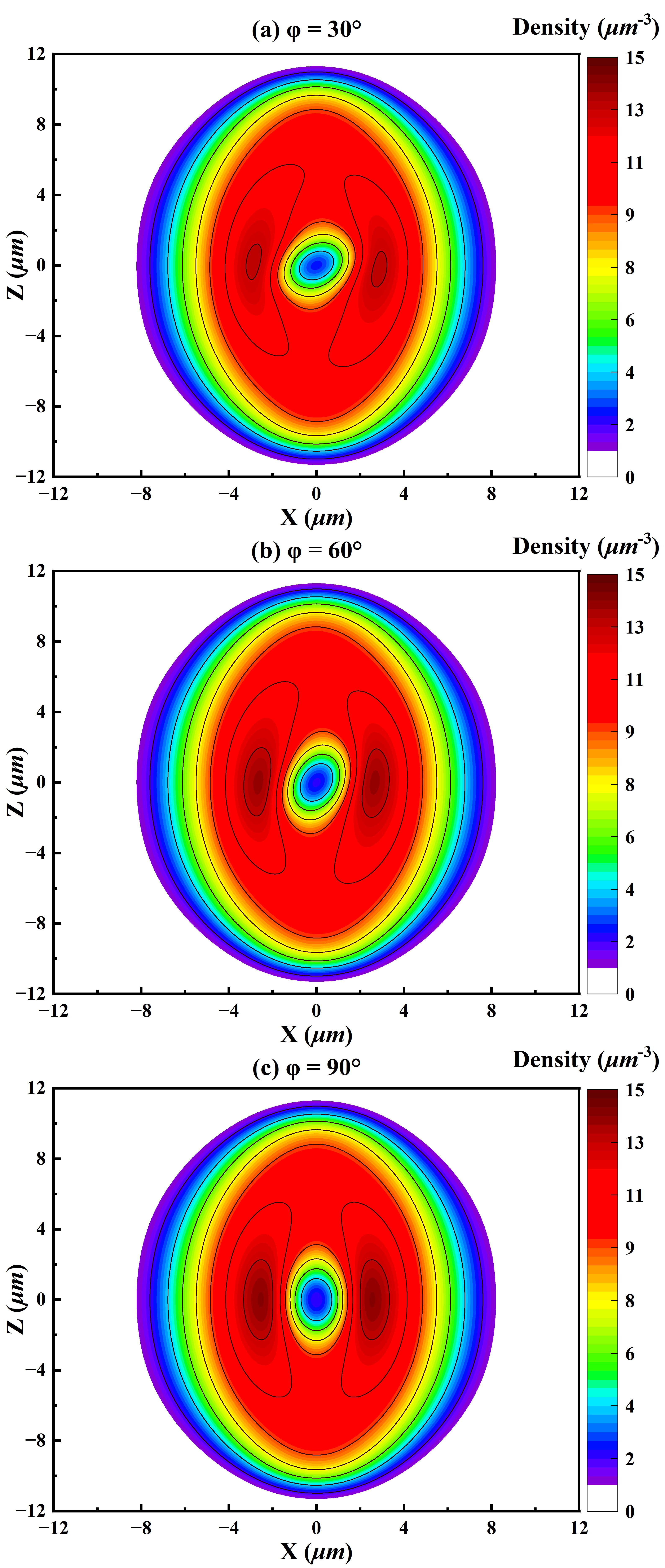}
    \caption{
    Density of the system (BEC + static impurity) in the $xz$-plane for different orientations of the impurity, i.e., for different values of $\varphi$, cf. Eq.~(\ref{eq:potential_rotated}). The deformation ratio is \(a=3b/2\) here.
    }
    \label{fig:3}
\end{figure}

We present densities for \(30^\circ\), \(60^\circ\), and \(90^\circ\) fixing $a=3b/2$ in Fig.~\ref{fig:3}. All these results exhibit similar features to those discussed in Fig.~\ref{fig:2}. Furthermore, they demonstrate the rotation of the density driven by the angle $\varphi$. For example, as the angle increases, the center appears to rotate and align with that particular angle.  The density contours almost appear 'stirred' though the impurity is static.  The contours in the case of the top two panels also show the gradual evolution from the shape and orientation of the impurity to that of the bath.  In the bottom panel, the impurity's elongation matches the elongation of the bath, so the center contours' orientation matches that of the edge contours.  In addition, the maximal density regions become more focused as the major axis of the impurity gets closer to aligning with the polarization direction.   

A superficial visual inspection of the density contours suggests that the maximum effect of the impurity potential occurs when the impurity's major axis is orthogonal to the polarization of the dipoles and decreases as they become more aligned.  This remains true even for larger deformations of the impurity (not shown). To quantify this observation, we calculate the energy of the system below.  

\begin{figure}
    \centering
    \includegraphics[width=0.5\textwidth]{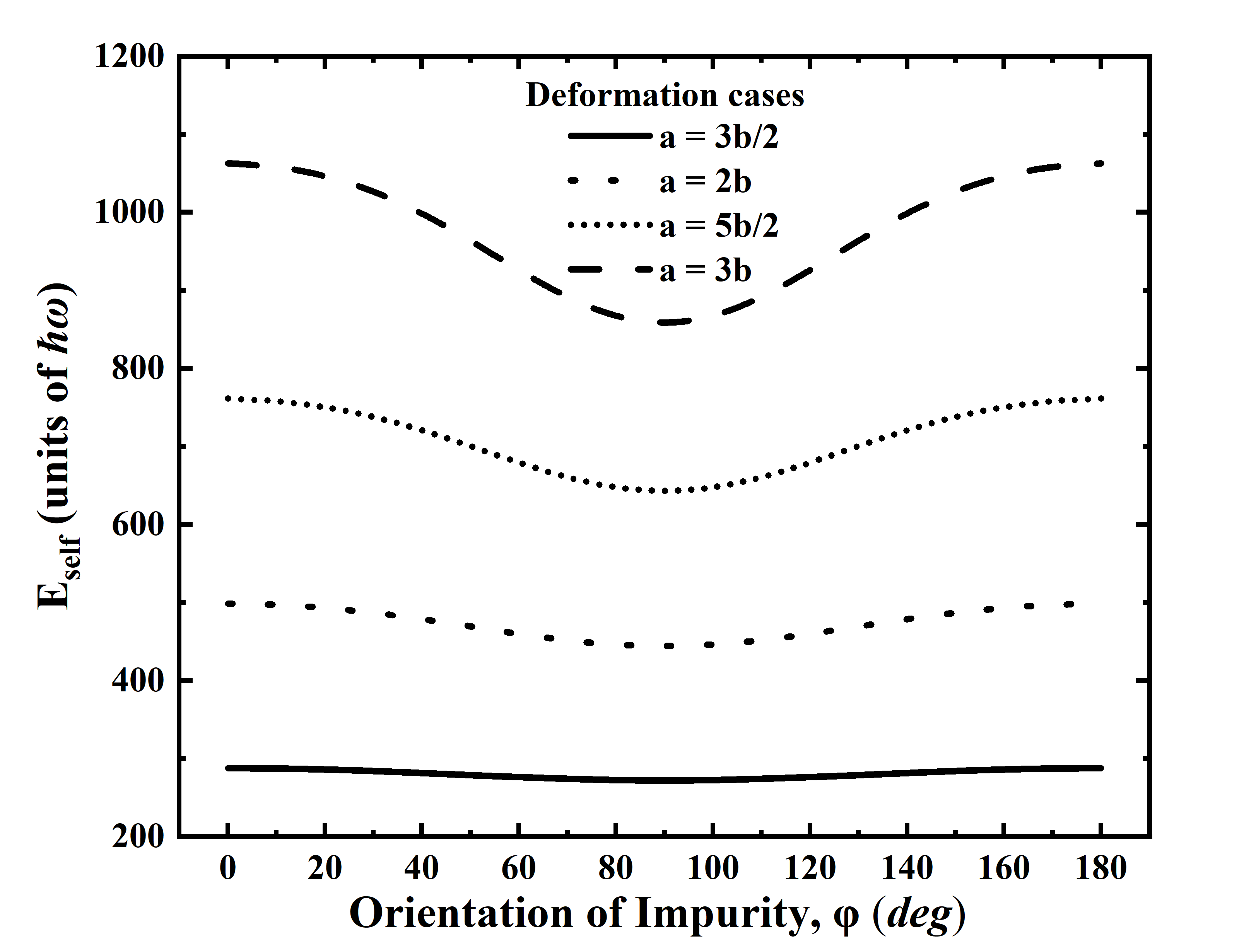}
    \caption{Effect of rotation on the self-energy of the impurity for different deformation ratios. Recall that the orientation angle is defined with respect to the $x$-axis and that the volume integral of the impurity ($\int\mathrm{d}\mathbf{r}V_i(\mathbf{r})= 3V_0L^3\sqrt{\pi}/4 $) is fixed, cf.~Eq.~(\ref{eq:potential_rotated}). 
    }
    \label{fig:4}
\end{figure}

\subsection{Self-Energy of  Impurity}
The results above show the interplay between the anisotropy of the impurity and the anisotropy of the medium in a trap. Quantifying this interplay by looking at the densities is difficult. Therefore, in this subsection, we study the energy of the system.  
 To quantify this effect, we calculate the self-energy ($E_{\mathrm{self}}$) defined as in Sec.~\ref{sec:homog_section}. We expect this quantity to depend not only on the amount of deformation but also on the orientation of the impurity.
 
 In Figure~\ref{fig:4}, the self-energy of the impurity for several deformation ratios is presented with respect to various orientation angles. The self-energy of the impurity-BEC system exhibits a nearly sinusoidal variation, showing a strong preference for the major axis of the impurity deformation to align along the \textit{z}-axis (cf. Eq.~(\ref{eq:self_energy_TF})).  The effect is solely due to the dipolar nature of the condensate, and vanishes if $\varepsilon_{dd}=0$.   Larger deformation ratios exhibit very similar qualitative behavior, though quantitatively all the self-energies increase with increasing deformation.

\begin{figure}
    \centering
    \includegraphics[width=0.5\textwidth]{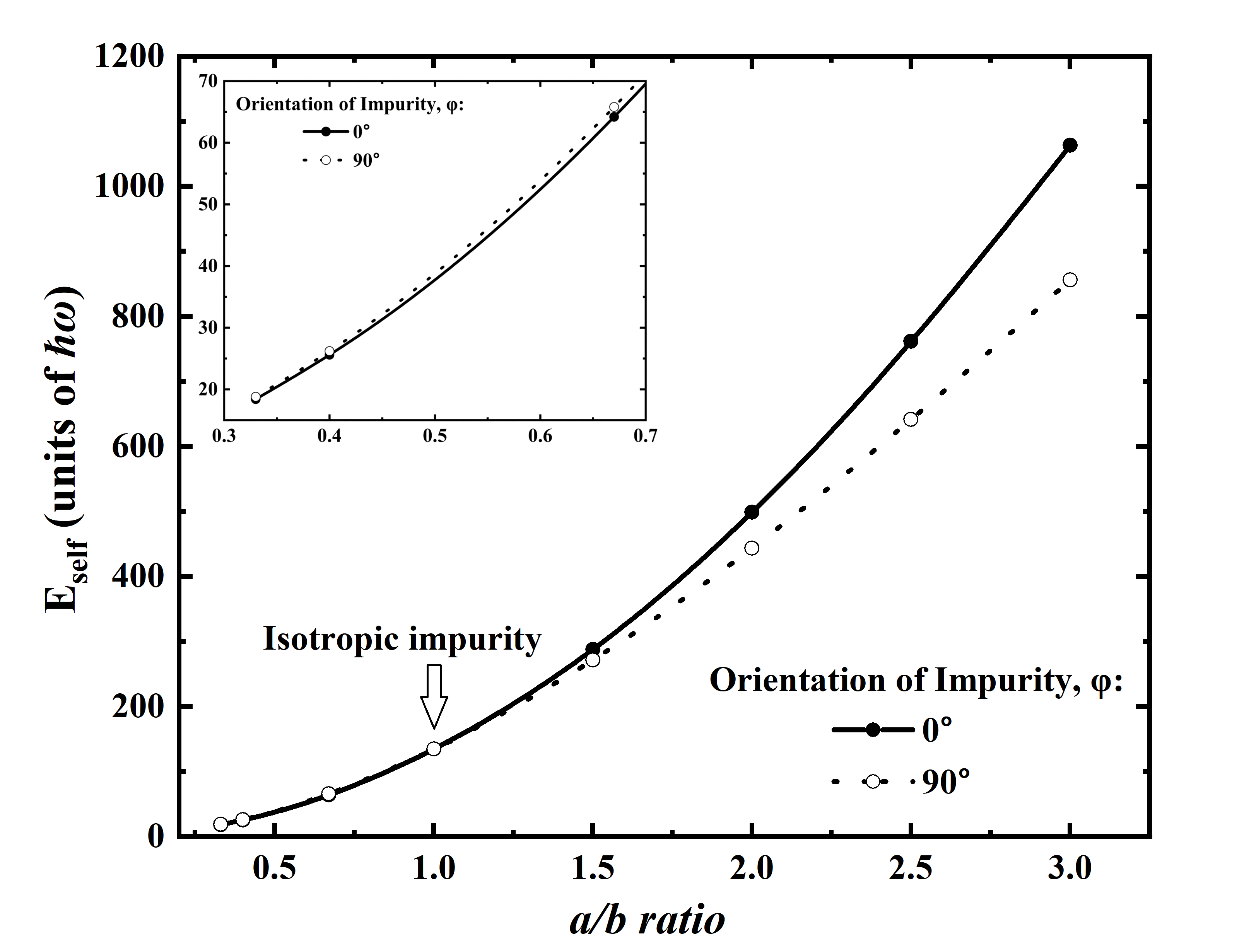}
    \caption{Self-energies plotted as a function of the deformation ratio, $a/b$, for two orientations of the impurity potential in Eq.~(\ref{eq:potential_rotated}). The inset zooms in on small $a/b$ ratios.}
    \label{fig:5}
\end{figure}

We illustrate this in Figure ~\ref{fig:5}, which demonstrates the dependence of the self-energy on the major and minor axis ratio, $a/b$.  This plot echoes what was seen in Figure~\ref{fig:4}, in that the most energetically favorable impurity is the one with the most elongation along the polarization axis.  The prolate ($a<b$) impurity allows more of the dipoles to be closer together in their favored head-to-tail configuration.  Once the impurity becomes isotropic and then oblate ($a>b$), the self-energy increases more rapidly as the impurity's presence is increasingly disruptive to the dipolar gas. The dashed line in Figure \ref{fig:5} shows the self-energy results for the systems where the impurity has been rotated by 90$^\circ$.  Here we see that for the oblate case, rotating the impurity decreases the self-energy.  Since after rotating 90$^\circ$, one of the impurity's major axes is now along the $z$-axis, this result is not surprising.  In the case of the prolate impurity, we observe the opposite in that the rotation increases the self-energy.  The reasoning is the same since now the longest axis is along the $x$-axis, though the difference is less dramatic than in the oblate case. 

\subsection{Time Dynamics}
\begin{figure}
    \centering
    \includegraphics[width=0.4\textwidth]{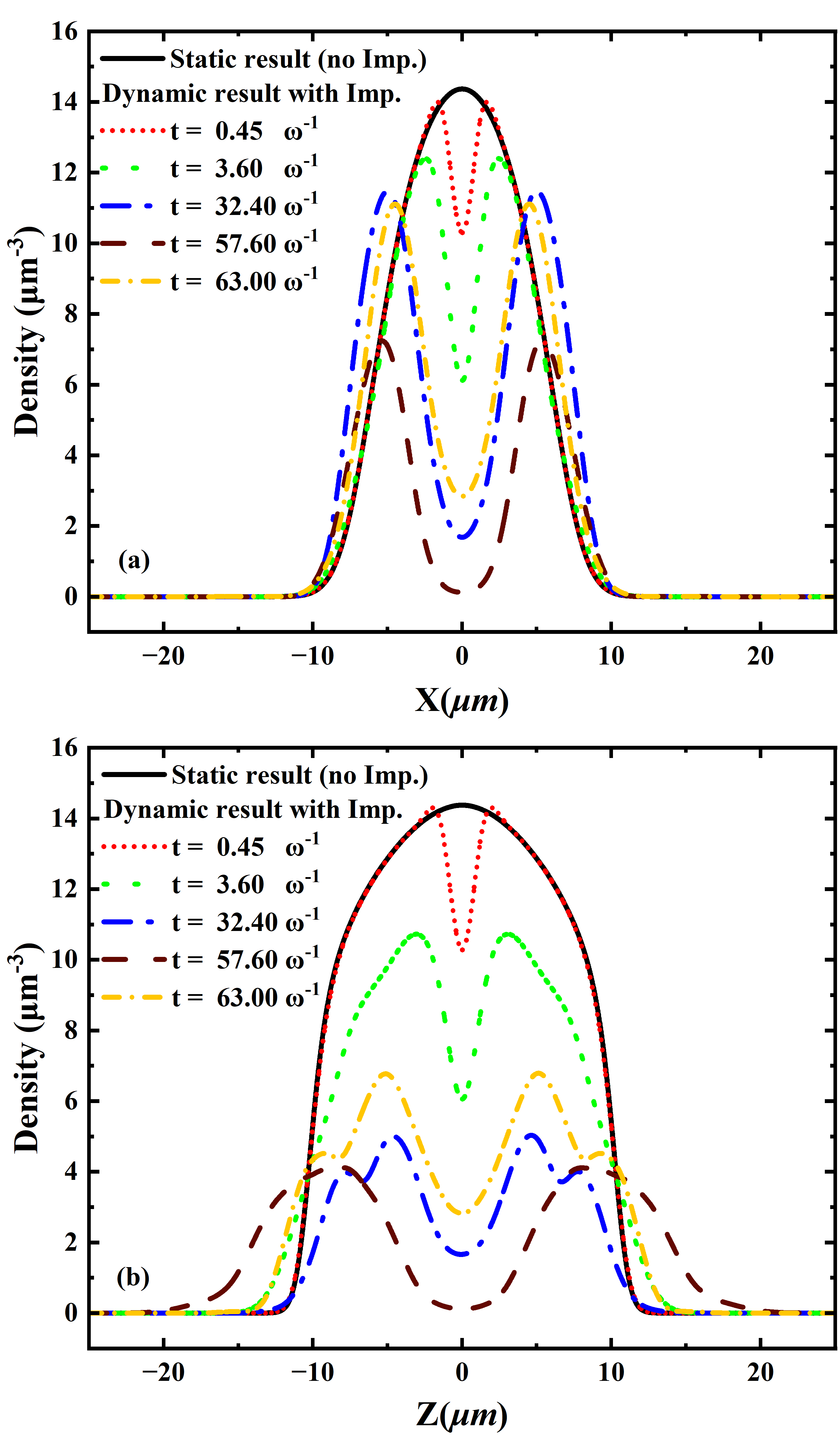}
    \caption{
   Density profiles along the $x-$axis (upper panel) and $z-$axis (lower panel) for different amounts of time evolution.  At $t=0$, a dipolar BEC without an impurity has a spherical impurity implanted into it at the origin. The center density decreases until t=57.60 $\omega^{-1}$, then begins to rebound. Note that the important timescale is defined by the frequency of the external trap, $\omega$, which sets the smallest energy scale of the system. For the Dy system, which we use in our numerical simulations, $1/\omega\simeq 2.5$ ms.}
    \label{Fig7}
\end{figure}

Another interesting question is how the system evolves once an impurity is introduced (going from Fig.~\ref{fig:1} to Fig.~\ref{fig:2}).  To study it, we start with the density solution of our system without an impurity and then utilize the real-time propagation of the split-step Crank-Nicolson method for our numerical solutions.  The results of this can be seen in Fig.~\ref{Fig7}, where the density in 1D cross sections along the $x-$ and $z$-axes are shown for an isotropic impurity.  Results for the other impurities were similar to the isotropic case, so we discuss the time evolution without focusing on the different impurities.  In Fig.~\ref{Fig7} it can be seen how the density plunges in the center of the trap by first creating large peaks close to the origin, which then recede over longer times as more particles are pushed away from the repulsive impurity and vacate the center region of the trap. 
In both plotted directions, the density plunges all the way to zero in the center before rebounding back,  and then repeating.  The repeating pattern is due to the trap and was also seen in non-dipolar trapped systems, see, e.g., Ref.~\cite{Akram2016}.  
\\
\indent An additional feature can be seen in the plots in the $z$-direction.
One can see a shoulder in the density peaks forming which separates into peaks seeming to show waves travelling outward along the $\pm z$-axes.  
This behavior is similar to what happens in a homogeneous non-dipolar case, see, e.g., Ref.~\cite{Marchukov2021}, with a crucial difference that the dynamics in the $x$ direction are different from that in the $z$ direction.
These waves appear to hit the trap boundary and then rebound and collapse eventually into a single peak as time increases.  Recall that the spatial extent in the $x$-direction (see Fig.~\ref{fig:1}) is not as large as that in the $z$-direction, and thus the impurity induces this behavior only in the $z$-direction.  Finally, these results were obtained within the mean field GPE framework, and further study involving beyond mean field effects and/or finite temperature could be explored in further study.

\section{Conclusion and Outlook}
\label{sec:conclusion}
We have investigated the response of a dipolar medium to an implantation of a repulsive impurity at its center.  We have observed a large distortion to the density of the medium, which indicates a preference for the dipoles to stay in their preferred head-to-tail configuration.  This preference is reinforced by our results for self-energy, which we have examined as a function of the deformation of the impurity as well as the orientation of the impurity.  Future experiments with impurities in dipolar BECs would expect to see anisotropic density modulations in their atom-cloud images regardless of the type of impurity used.  Mobile impurities, though smaller in size than the impurity potential of this work, should also lead to this behavior. It might be however harder to study the density modulations in this case as mobile impurities are free to move within the Bose gas distorting the signal.
Our results pave the way for a number of possible follow-up studies some of which we briefly outline below. 

{\it Mixtures.} Mixtures of dipolar Bose gases with non-dipolar Bose or Fermi gases received relatively little attention in spite of their potentially rich physics~\cite{Dutta2010,Kain2011}.
Recent progress in realizing quantum mixtures of dipolar gases~\cite{Trautmann2018} may, however, motivate further exploration of such systems.

Inadvertently, our results provide insight into properties of two-component mixtures. 
Arguably, the simplest mixture  is an impurity system --  a dipolar polaron~\cite{Kain2014,Ardila2018,Volosniev2023} -- that was mentioned already in Sec.~\ref{sec:homog_section}. The self-energy in Eq.~(\ref{eq:self_energy_TF}) suggests that the induced impurity-impurity interaction becomes long-range. Indeed, assuming that the $V_i=v_1+v_2$, where $v_1$ and $v_2$ correspond to two  impurities, we derive the dipole-interaction-induced part of the mediated  interaction as
\begin{equation} 
V_{\mathrm{eff}}\simeq \frac{1}{g^2}\int v_1(\mathbf{r})V_{d}(\mathbf{x}-\mathbf{r})v_2(\mathbf{x})\mathrm{d}\mathbf{r}\mathrm{d}\mathbf{x},
\end{equation}
which, assuming that $v_j(\mathbf{r})=g_j\delta(\mathbf{r}-\mathbf{a}_j)$, leads to $V_{\mathrm{eff}}\simeq g_1g_2 V_d(\mathbf{a}_1-\mathbf{a}_2)/g^2$. This interaction is of long-range in contrast to the well-known mean-field effective impurity-impurity interactions, see, e.g.,~\cite{Bruderer_2008,Naidon2018,Brauneis2021,Petkovic2022,Will2021}. In the future it will be interesting to investigate this potential in the vicinity of the collapse where the impurities strongly modify the bath, leading to strong mediated correlations.

 Finally, we note that one can naively think of the impurity potential in Eq.~(\ref{ImpEqn}) as of some zero-range-interacting Bose or Fermi gas that does not mix with the dipolar BEC, for conditions of miscibility of Bose-Bose mixtures see Ref.~\cite{Kumar_2017}. In this case, our study provides some ideas for understanding immiscible two-component gases, where one of the gases (impurity) has only zero-range interactions.  
 For example,  Figure~\ref{fig:5} suggests that the density of this impurity gas will be shaped by the dipolar interaction.

\section*{Acknowledgements}
The authors acknowledge that this material is based upon work supported by the National Science Foundation/EPSCoR RII Track-1: Emergent Quantum Materials and Technologies (EQUATE), Award OIA-2044049.

\appendix

\section{Derivation of Eqs.~(\ref{eq:self_TF}) and~(\ref{deltaN})}
\label{app:Eq13}

First, let us outline a derivation of Eq.~(\ref{eq:self_energy_general}). We note that the energy within the mean-field approximation is given by
\begin{align}
 E=&\mu N -\frac{gN^2}{2}\int\psi(\mathbf{x})^4\mathrm{d}\mathbf{x}- \nonumber \\
 &\frac{N^2}{2}\int\psi(\mathbf{x})^2V_d(\mathbf{r}-\mathbf{x})\psi(\mathbf{r})^2\mathrm{d}\mathbf{r}\mathrm{d}\mathbf{x},
 \label{eq:app_energy}
\end{align}
where $\psi$ is the solution of the corresponding Gross-Pitaevskii equation. Without the impurity $\psi(x)$ is constant, whereas in the presence of the impurity it depends on the position. Equation~(\ref{eq:app_energy}) together with Eq.~(\ref{eq:self_energy_def}) leads to Eq.~(\ref{eq:self_energy_general}). To derive Eq.~(\ref{eq:self_TF}), we  
use Eq.~(\ref{eq:GPE_Thomas_Fermi}) together with the identity
$\delta n(\mathbf{x})=N|\psi(\mathbf{x})|^2-n_0$ in Eq.~(\ref{eq:self_energy_general}).

We derive Eq.~(\ref{deltaN}) using Eq.~(\ref{eq7}). 
Indeed, by definition, 
\begin{equation}
\delta N=-\frac{1}{g(2\pi)^3}\int\mathrm{d}\mathbf{k}\mathrm{d}\mathbf{r} \frac{\tilde V_i(\mathbf{k})}{1+2\varepsilon_{dd} P_2(\cos\alpha)} e^{i\mathbf{k}\cdot\mathbf{r}}.
\end{equation}
Noticing that $\delta(\mathbf{k})=\frac{1}{(2\pi)^3}\int\mathrm{d}\mathbf{r}e^{i\mathbf{k}\cdot\mathbf{r}}$, we rewrite this expression as 
\begin{equation}
\delta N=-\frac{\tilde V_i(0)}{g}\int\mathrm{d}\mathbf{k} \frac{\delta(\mathbf{k})}{1+2\varepsilon_{dd} P_2(\cos\alpha)},
\end{equation}
where we assume that $\tilde V_i(0)$ is well-defined. As the delta function is spherically symmetric, we can first take the integral over the $\alpha$-angle
\begin{equation}
\int_{-1}^1 \frac{\mathrm{d}x}{1+2\varepsilon_{dd} P_2(x)}=\frac{2 \tan^{-1}\left[\sqrt{\frac{3 \varepsilon_{dd}}{1-\varepsilon_{dd}}}\right]}{\sqrt{\varepsilon_{dd}(1-\varepsilon_{dd})}}.
\end{equation}
Utilizing this expression, the fact that $\int k^2\mathrm{d}k \delta(\mathbf{k})=1/(4\pi)$ and $\tilde V_i(0)=\int \mathrm{d}\mathbf{r}V_i(\mathbf{r})$, we derive the expression presented in Eq.~(\ref{deltaN}).

\section{Attractive Impurity}
\label{app:attractive}

\begin{figure}
    \centering
    \includegraphics[width=0.5\textwidth]{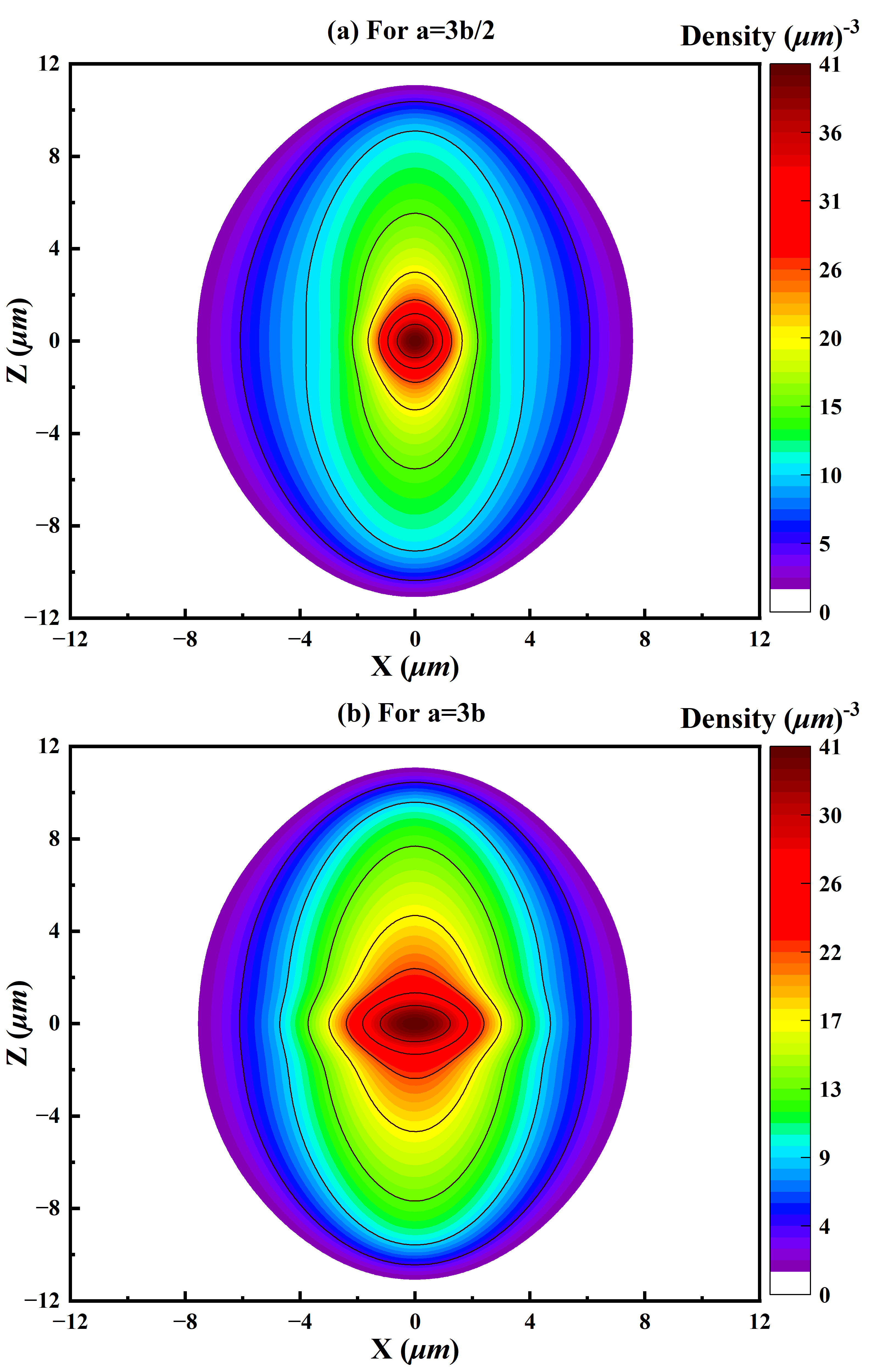}
    \caption{Density contours in the $xz$-plane for an attractive Gaussian impurity implanted into a dipolar BEC for two different deformation ratios $a/b$.  The dipoles are polarized in the $z$-direction. }
    \label{fig:appendix-figure2}
\end{figure}

For an attractive impurity $(V_0 < 0)$, particles will naturally be attracted to the center of the trap.  We demonstrate contour plots for attractive impurities in Fig.~\ref{fig:appendix-figure2} showing both an isotropic impurity and our extreme deformation, $a=3b$.  One can see the very large density peak in the center, which is over three times as dense as the maximum densities seen in the repulsive case.  Like before, the system is three dimensional so this is an ellipsoid of maximum density in the center of the trap.   Interestingly, just as in the repulsive plot, there appears to be a slight bump in the contour line as it crosses the $x$-axis (at around $\pm$ 4 $\mu$m) for the largest non-spherical case, though the bump goes in the opposite direction.  In both the attractive and repulsive cases it makes sense.  In the attractive case it goes outward indicating a slight bulge of larger density created by the attractive impurity, and it is the opposite in the case of repulsion.

\begin{figure}
    \centering
    \includegraphics[width=0.5\textwidth]{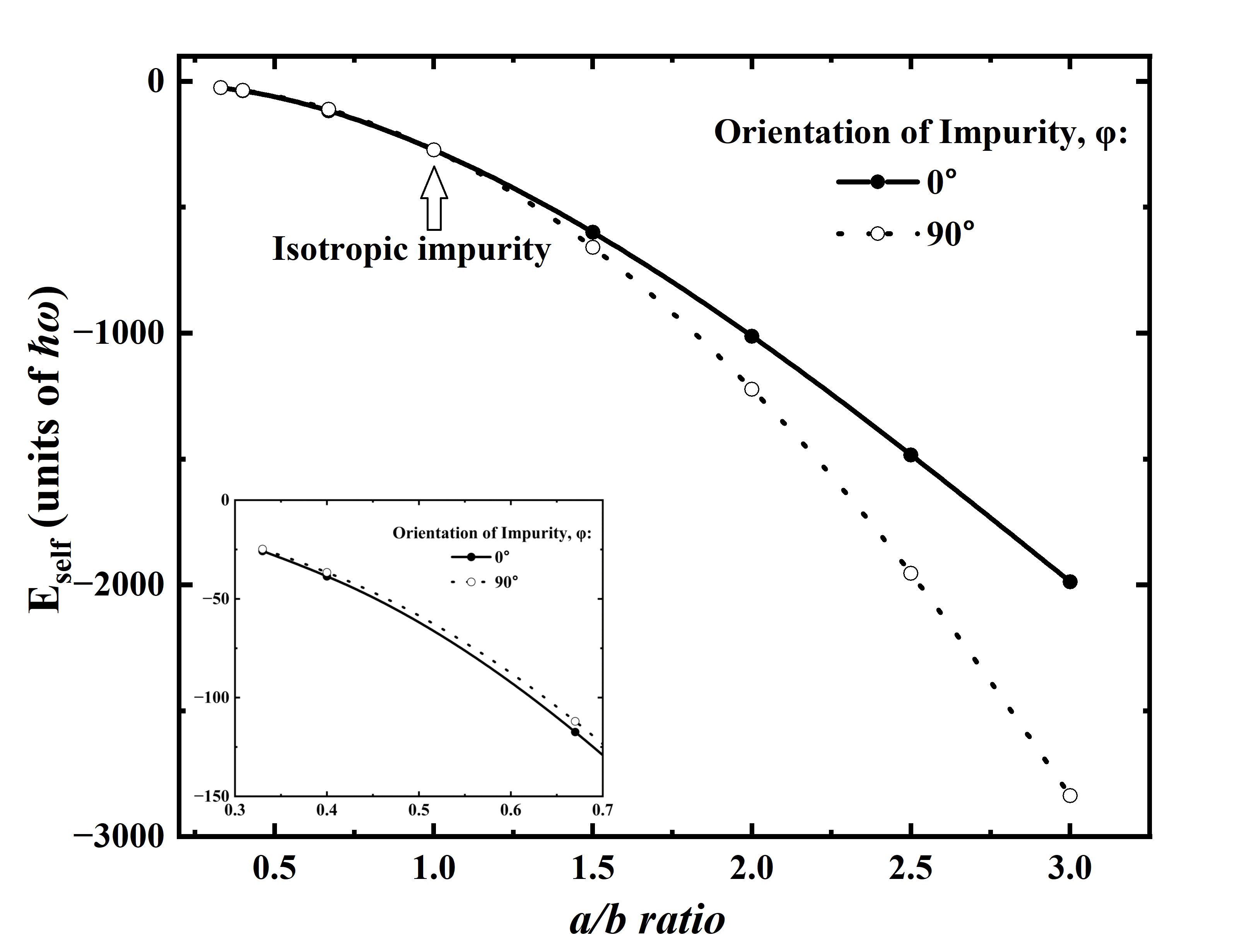}
    \caption{This figure shows the dependence of the self-energy as a function of the deformation of the impurity potential (expressed as the ratio $a/b$) for an attractive impurity. The inset displays a close-up for small $a/b$ ratios. }
    \label{fig:appendix-figure1}
\end{figure}

In Fig.~\ref{fig:appendix-figure1}, we show the self-energy as a function of deformation plot analagous to Fig.~\ref{fig:5}.  Here we see that, qualitatively, the plot looks the same except it is reflected over the $x$-axis.  The relationship between the $0^\circ$  and $90^\circ$ curves, however, remains the same with the rotated impurities, where the rotated curve is lower in energy for oblate impurities and higher for prolate ones.  The magnitudes are also about twice that of the repulsive case for the solid curve.  It is not surprising that adding any kind of attraction lowers the energy of the system, but this figure shows that an oblate impurity with one of its major axes along the polarization direction of the dipoles is the most stable.   In the limit where the impurity becomes a disc, then it would match the shape of the the contours in the top panel of Fig.~\ref{fig:1}, and thus matching the shape of the dipolar gas.     

\bibliographystyle{apsrev4-1}
\bibliography{ref}



\end{document}